# Energy Systems Test Case Discovery Enabled by Test Case Profile and Repository

Petra Raussi
*Carbon Neutral Solutions*
VTT Techn. Research Centre of Finland
Espoo, Finland
0000-0001-9768-8312
petra.raussi@vtt.fi

Jirapa Kamsamrong
*Energy Division*
OFFIS e.V.
Oldenburg, Germany
jirapa.kamsamrong@offis.de

Alexandros Paspatis
*School of Electrical and Computer Eng.*
National Technical University of Athens
Athens, Greece
0000-0002-3479-019X
agpaspatis@mail.ntua.gr

Kai Heussen
*Wind and Energy Systems*
Technical University of Denmark
Roskilde, Denmark
0000-0003-3623-1372
kheu@dtu.dk

Tesfaye Amare Zerihun
*Energy Systems*
SINTEF Energy AS
Trondheim, Norway
0000-0002-3893-957X
tesfaye.zerihun@sintef.no

Edmund Widl
*Center for Energy*
AIT Austrian Institute of Technology
Vienna, Austria
0000-0002-2834-306X
edmund.widl@ait.ac.at

Filip Pröstl Andrén
*Center for Energy*
AIT Austrian Institute of Technology
Vienna, Austria
filip.proestl-andren@ait.ac.at

Jawad H. Kazmi
*Center for Energy*
AIT Austrian Institute of Technology
Vienna, Austria
0000-0001-9720-6862
jawad.kazmi@ait.ac.at

Thomas I. Strasser
*Center for Energy*
AIT Austrian Institute of Technology
Vienna, Austria
0000-0002-6415-766X
thomas.strasser@ait.ac.at

Felipe Castro
*Engineering Services*
SMA Solar Technology AG
Kassel, Germany
fcastro@sma.de

Luigi Pellegrino
*Materials and Generation Technology*
Ricerca sul Sistema Energetico S.p.A.
Milan, Italy
0000-0002-1258-9516
ing.pellegrinoluigi@gmail.com

*Abstract*—Smart energy systems comprise multiple domains like power, thermal, control, information, and communication technology, which increases the complexity of research and development studies. This expansion also requires larger and ever so complex experimental pilot environments driving the demand for geographically distributed multi-research infrastructure tests. The Holistic Test Description approach supports the design of multi-domain and multi-research infrastructure tests by organizing the test cases into comprehensive segments, ensuring all relevant items for testing are covered. These test cases eventually form a pool, which to understand holistically would require studying and reading all the descriptions. This work proposes therefore the concept of Test Case Profiles to improve test case discovery and the structured creation of them. Test Case Profiles add further structure to the indexing in test case repositories. Along with the proposed indexing method, four different use cases are introduced to motivate additional applications of the proposed concept.

This work received funding in the European Community's Horizon 2020 Program (H2020/2014–2020) under project "ERIGrid 2.0" (Grant Agreement No. 870620).

## I. INTRODUCTION

Energy system modeling and solution validation are becoming increasingly complex. While previous studies could be very limited in scope and still provide valuable novelty, there is a need to find global optimal encompassing large multi-domain environments in the modern world. Therefore, smart energy system solutions should not be developed in isolation but exposed to other sectors, such as thermal and hydrogen systems. Sector integration increases the complexity of the Test Cases (TCs) and the difficulty in planning and designing tests due to the traditional disciplinary silos.

Before actual deployment, testing in a laboratory environment is quite often required to verify proof of concept. This necessitates a test setup that incorporates hardware and software to emulate a real-world testing environment such as Software-In-the-Loop (SIL) and/or Hardware-in-the-Loop (HIL) in form of Controller Hardware-in-the-Loop (CHIL) or Power Hardware-in-the-Loop (PHIL) [1]–[4]. However, it is not easy to design multi-domain tests and execute them.

The equipment required from each domain might not be abundantly available at each Research Infrastructure (RI). Therefore, multi-RI experiments are encouraged, in which the System under Test (SuT) and the test equipment are divided among several RIs, in different geographical locations. There are various experiment types in which two or more RIs can be interconnected [5]. Executing such experiments among several RIs and interconnection platforms deepens the complexity, further motivating the use of structured test planning and execution. In [6], a structured method, Holistic Test Description (HTD), is introduced to support the design of multi-domain TCs. The HTD organizes a test case into comprehensive segments ensuring that all relevant items for the development are considered.

A proven strategy to improve the quality of TC descriptions is the provision of exemplary TC formulations, as is practiced with Use Case (UC) repositories [7], [8]. In this way, a collection of related TCs could be employed as a structured creation of TCs. However, it can be challenging to gain a quick overview of a pool of TCs while generating a multifaceted collection of TCs on the same research problem from various aspects. This motivates the question of how similarity between TCs may be defined – since a pure index-based search may prove inaccurate due to the multi-perspective nature of TCs.

This work therefore proposes the concept of Test Case Profile (TCP) to improve TC discovery and structured creation of them. TCP adds a further structure to the indexing in TC repositories. Along with the proposed indexing method, four different UCs motivate additional applications of the proposed concept. The TCP approach is similar in principle to other multi-dimensional keyword identification approaches known from the literature [9].

The remaining parts of the work are organized as follows: Section II provides the definition of TCPs and related context information whereas the formulation of them is presented in Section III. The usage of the TCPs on selected four UCs is shown in Section IV followed by a discussion and reflection in Section V. Finally, the conclusions and main findings are summarized in Section VI.

## II. TEST CASE PROFILES DEFINITION AND CONTEXT

In this section, the proposed concept of TCPs is first defined, then the background for the definition and use is provided and finally, the basic usage of the TCPs is described.

### A. Test Case Profile Concept Definition

A TCP describes a collection of TCs that share similarities, both in the context of application and testing facility properties. In contrast to common methods for describing and indexing UCs, the four dimensions of the TCP cover *Domain under Investigation*, *Tested phenomenon*, *Type of Assessment*, and *Test System/Components*. Associated with each of these dimensions is a defined set of keywords [10].

The background for indexing in these four different perspectives rests in the structure of test descriptions, multi-domain testing problems, and associated experimental setups, as summarized in the following. Conceptually, it is worth noting that Use Case (UC) is not among the TCP dimensions. This highlights the point that TCs and test infrastructure span across ranges of applications and associated UCs.

### B. Test Descriptions Structure

The HTD approach, outlined in [6], proposes organizing a TC's high-level description into structural, functional, and testing purpose elements. The test description further comprises refinements (mapping steps): *(i)* the test specification (model structure, procedure, variables, and functionals) and *(ii)* the experiment specification (the mapping to a RI). Moreover, an extension of the HTD approach was presented in [11]. Building on the HTD description concepts, it extends the high-level scope with a test with a phenomenon under test.

### C. Test Descriptions Organisation "Functional Scenarios"

Various topics and themes in smart energy systems create a convoluted environment to navigate. An approach to structure high-level visions is the Functional Scenario (FS), presented in [12], which is an umbrella term comprising a strategic high-level vision for RI applications. FS includes subsections in an increasing granularity and is one way to form a group of TCs that derive from the same motivation. FS targets to support research and development work by providing a high-level perspective of the purpose of the work.

A FS consists of System Description, Motivation, Use Case, Test Case, Experimental Setup, and Relevance Sections as depicted in Figure 1. The System Description depicts the physical system and environment in which the FS is situated. Motivation and Relevance describe why this particular FS should exist to address a specific challenge. Use Case, Test Case, and Experimental Setup work in steps of granularity, with the Use Case explaining the system's behavior, the Test Case how this could be tested, and the Experimental Setup of the type of devices and components that would be required to execute the test [12].

Fig. 1. Overview of a Functional Scenario structure [12].

### D. Indexing Test Descriptions with Test Case Profiles

The test descriptions are indexed using predefined keywords in each of the four TCP dimensions. For instance, the nominal behavior of the system which is being tested can be vastly different in TCs within the TCP. The TCs in a TCP could be, for example, applying the same testing methodology to their respective UCs. The UC defines the behavior or interaction of components and a TC specifies the required information of the components, i.e., test setup [13]. TCP is a tool to harmonize and structure TCs into comprehensible collections.

The main focus is on practicality and usability for research rather than high-level visionary strategy work. An application of TCP for this purpose is reported in [10], where the technical content of test cases associated with FSs is summarised by means of TCPs.

## III. FORMULATING TEST CASE PROFILES

Smart energy systems usually deal with multi-domains which necessitate multidimensional tests. Formulating a TC is motivated by objective, method, and system, which helps users to identify a high-level TC narrative. To identify a TC in detail, users can specify technological areas like power system, multi-domain energy system, automation and control as well as Information and Communication Technology (ICT) integration. Keywords could be used to formulate the TCPs related to intended testing.

In the H2020 ERIGrid 2.0 project [14], twenty-five TCs are developed and assigned by keywords that represent a characteristic of a specific TCP [10]. The keywords assigned aim to identify fundamental concepts so that users with little prior knowledge of the TCs could be able to shortlist a selection of keywords associated with a given TC.

The keywords are identified within each of the four dimensions that have been identified as critical to identify a testing context: *Domain under Investigation*, *Phenomenon under Test*, *Type of Assessment*, and *Test System/Components*, as shown in Figure 2. Domain under Investigation refers to a larger domain, the TC focusing on highly multi-domain TCs and providing an easy-to-understand first impression of what field(s) the TC is targeting. The Phenomenon under Test describes the behavior of the entity being tested, while the Type of Assessment refers to the research methodology used. Since some of the keywords are fairly common, there could be a danger of keywords possessing several meanings, especially in the context of multi-domain experiments; therefore each of the keywords has been provided with definitions [10].

There are diverse user groups in smart energy systems research. Given that TCP users vary depending on their knowledge and motivation, a keyword approach provides a foundation for developing an appropriate TCP for each user group. In Section IV this work presents four UCs to reflect practical examples and cover relevant user groups varying from basic to advanced users, but not limited to managerial users. TCP users should have a fundamental knowledge of how to formulate and/or filter a relevant TCP. Later on, an experienced user can use TCP to identify research gaps within a certain technological domain, such as developing a reference benchmark for multiple TCs. A research project involves multiple stakeholders from various backgrounds; stakeholders can use TCP to develop coherent test cases to help the project achieve its aims. Furthermore, TCP can be used by managerial users to examine the possibilities and future research and development direction considering RIs collaboration.

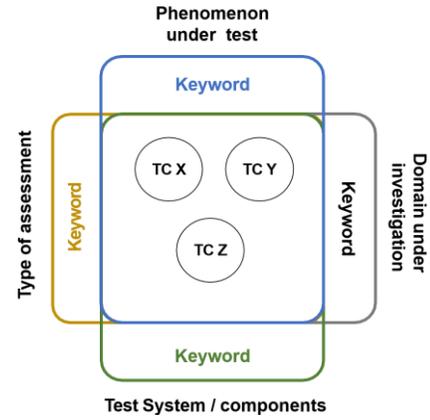

Fig. 2. Visual illustration of a Test Case Profile set of Test Cases with keywords in four dimensions [10].

The keyword had to be identified following the categorical dimensions listed above (i.e., *Domain under Investigation*, *Phenomenon under Test*, *Type of Assessment*, and *Test System/Components*). Within each dimension, suitable keywords had to be identified, balancing abstraction with sufficiently concrete technical areas. For the H2020 ERIGrid 2.0 project, the selected keywords are shown in Figure 3.

## IV. USAGE OF TEST CASE PROFILES

TCP users have diverse backgrounds and interests. Beginner users often face finding a reference TC that can be applied to their TC, while experienced researchers are searching for a suitable benchmark for a specific test and harmonizing several TCs to achieve a project goal. A program manager is one user group that can make use of TCPs for providing laboratory tests and collaboration between RIs. This section presents four UCs for the application of TCP, each associated with one of these user groups.

### A. Beginner Entering to New Domain

Interdependence between systems requires tests dealing with complex experiments and interactions between domains. Students and junior researchers with pure individual backgrounds may not have experience in other domains, which can cause challenges in formulating the TC. For example, a student with a pure electrical engineering background may not be familiar with ICT interaction for providing observability and controllability in smart energy systems. This user group lacks experience in some domains, how to investigate, what system to consider, and how to minimize foreseeable uncertainties for their tests. In this case, TCP can provide an example that can guide them for methods, types of assessment, and components of a certain test phenomenon by considering correlation across different domains.

Users should first identify the narrative test objective and why this TC is investigated, i.e., the Phenomenon under Test.

Fig. 3. Identified keywords for profiling multi-domain energy system Test Cases in the H2020 ERIGrid 2.0 project [10].

Keywords can be used to identify the characteristic of the TC. To clarify the UC, an example of the interdependency of communication components in the power system is represented for a multi-domain TC. This example is to identify the communication packet loss of Intelligent Electronic Devices (IEDs) that would affect control command responsiveness which could help a user unfamiliar with communication performance testing. In this example, a user can first select the keywords *Control* and *ICT* in the Domain under Investigation dimension and *IEDs* for test components by using TC profile keywords [10], [15]. Later, the keyword *Packet Loss* in the phenomenon under test dimension can be used to investigate the characteristic of communication delay. As a result, a set of Test Cases of TC17, TC23, TC24, and TC25 are selected regarding selective keywords in the H2020 ERIGrid 2.0 projept [15].

At this stage, a list of TCs is provided for a user to identify further the reference TCs that may be suitable for their experiment. If the user aims to study the impact of ICT on a different domain, additional keywords could be used to narrow down the list of TCs. For example, a user wants to investigate the impact of packet loss on the power system operation. The use of keyword *Energy Balance* in the test phenomena and keyword *Communication Performance* will result in only one TC which is TC24, as shown in Figure 4. At this point, a user can use this reference TCs to formulate their test design relevant for testing the impact of communication delay of IEDs on a power system operation. It should be noted that this is a selective example to reflect how to use keywords for beginners which can be further applied in multi-domain UCs.

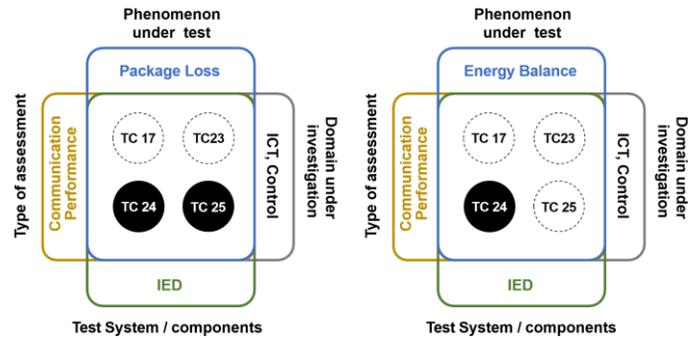

Fig. 4. Exemplary formulating Test Case Profile for a beginner [10].

### B. Suitable Benchmark Identification

An interesting problem arises regarding selecting the appropriate benchmark network when aiming to validate a series of TCs. In this case, the case study may be predefined based on some contractual requirements or research interests of the user, but the network to perform those studies is missing. In this approach, HTD is proposed as a tool to document all the required case studies, naturally leading to the specifications that such a benchmark system should have. Notably, this approach has been the main idea behind benchmark electricity networks, as they could be used while testing different methodologies and algorithms, thus enhancing the capability for a straightforward replication of results and further comparison, highlighting the novelties that new technologies bring. The HTD-based approach for developing benchmark networks has analytically been presented in [16], [17].

This UC is especially relevant for research groups when formulating further research. A research group can have the same high-level objectives and targets to add to their existing research and capabilities so that their research would converse to increase the group's overall level. Funding for research is often sourced from various types of research projects. Each project has its contracts and agreements on what should be accomplished. By deriving keywords for the TCPs from project agreements and building a TC collection, a group of case studies can be tested under the same network configuration. Furthermore, this approach could be used to develop a benchmark network to perform all the required studies as per the project agreements.

### C. Project's Case Studies Presentation

In this case, multiple stakeholders need to perform case studies on topics close to each other; however, their development and presentation may be frustrating with each stakeholder working in silos. To address this, HTD can be utilized so that the presentation can be unified. Such an approach has been followed in the stability studies of the different demo microgrids like in the H2020 RE-EMPOWERED [18] and ERIGrid 2.0 [10] projects.

Project consortia and industrial players can use this approach to form a coherent collection of TCs and avoid any logical gaps in the research, leading to more efficient collaboration. In fact, by checking the TCs with a tool like TCP, any silos or gaps between the TCs can be identified and addressed based on the keywords as for example shown in Figure 5. With this approach, the case studies within a project could be ensured to form a coherent narrative. The main aim of this UCs is to form a homogeneous presentation of various case studies bringing clarity to the presentation of the collection of case studies from multiple stakeholders.

Fig. 5. Illustration of applying the Test Case Profile keywords of a set of Test Cases belonging to one Functional Scenario [10].

### D. Aligning Infrastructure Capabilities with Roadmaps

Test descriptions and TCs are a vehicle to formulate requirements to RIs. As a TC represents a connection of a given solution/technology with a test-specific Technology Readiness Level (TRL) level, a TCP establishes this linkage of a field or area of technology and its maturity with suitable RI requirements. A program manager would maintain an overview of technologies and their maturity. On the other hand, a laboratory manager would maintain an overview of the capabilities of the managed RI. The TCP helps identify sets of requirements for a laboratory's capability development by identifying relevant TCs.

It could be difficult to describe the work between a laboratory manager and a program manager since there is a difference in required granularity. TCPs could help bring forward ideas for testing from all laboratory members neatly documented based on the HTD templates and indexed by the keywords. Thus the program manager could receive a holistic overview of the TCs as TCPs with the keywords providing the necessary level of detail and still have the TCPs attached to the detailed descriptions of each TC. By the identified TCP, it could be easier to build a classification of ongoing activities and laboratory capabilities. This UC could also support mapping research and development activities among several domains.

## V. DISCUSSION AND REFLECTION

Four UCs for applications of the concept of TCPs have been provided. Three of them are mainly for existing pools of TCs or TCPs; one is to form new TCPs based on contractual criteria. The first UC picks a new domain of interest based on keywords, studies the TCs matching with the domain, and then narrows down the number of TCs by picking further keywords from the other four dimensions. The second UC aims to develop a TCP by selecting suitable keywords based on project contract requirements and building detailed TC descriptions around these keywords. The third UC assumes that the collection of TCs is already indexed with keywords and forms coherent and harmonized TCPs and corresponding narratives based on which keywords have been selected. The fourth UC focuses on the selected keywords within the TCPs to form a holistic understanding and overview of the TCs.

Collections of TCs could be even more beneficial when the collection approach is structured and well organized. Thus, it will be possible to form a holistic overview of the TCs, without studying them in detail, but enabling the detailed study by including the full description per the HTD approach. With a combination of HTD-based TC descriptions and TCPs, it is possible to study the TCs from various levels of granularity.

Thanks to the varied granularity, the potential UCs of TCPs can be on several stations, from new researchers entering a field to research groups to higher-level program managers. Using a keyword-based approach that aims to remain highly simplistic, the adoption threshold is low as it is possible to study TCPs based on their keywords without prior knowledge and potentially use the available definition for the keywords as a guide to understanding the given TCP better.

Furthermore, the in-depth knowledge of the TCs is not lost in forming TCPs but remains at the core. Thus the integrity of the approach developed on top of the HTD is maintained. All the information is provided to the person studying the TCs through TCPs; at their discretion, they can read either the detailed descriptions or parts of them.

Within the context of the H2020 ERIGrid 2.0 project, working on the TCPs supported funneling the vast amount of TCs, which are all relevant for the project context, into a few concrete benchmarks and demonstrations still encompassing ideas from multiple TCs and having the possibility to further development to more detailed directions of any of the TCs. All TCs are open-source and freely available[1]. Other approaches for narrowing down to form the demonstrations could have been, for instance, voting among the consortium members, but by developing TCPs, the TCs naturally formed into logically structured and harmonized collections based on the keywords.

## VI. CONCLUSIONS

Multi-domain and multi-RI experiments are increasingly important due to sector integration and the drive of large-scale pilot environments. Both of these elements increase complexity in TCs and experiments. To capture all the relevant details of TCs into a logical format HTD approach has been used. The TCs described in detail form a pool, which to understand holistically would require studying and reading all the descriptions. To alleviate this challenge, the TCP approach based on keywords has been introduced in this work. With TCP, each TC is indexed with keywords from an existing predefined list of keywords and can be allocated to a collection of TCs which together form the profile. TCP provides structure and organization to pools of TCs containing various topics and experiment types.

---

[1] https://github.com/ERIGrid2/test-case

The keywords are allocated in four dimensions covering *Domain under Investigation*, *Tested Phenomenon*, *Type of Assessment*, and *Test System/Components*. The UCs for the TCPs range from new researchers and students entering a field to research groups and project consortia harmonizing their work to program managers forming a holistic overview of laboratory capabilities. Depending on the UC, existing keywords can be used to select TCs to study in detail, TC benchmarks formed based on keywords selected due to contractual reasons, or to form a holistic understanding of testing capabilities and research directions.

Finally, the future work will focus on the usage of the developed method in other research and development projects as well as the collection of additional TCs and TCPs.

## VII. Acknowledgment

The authors express thanks to all involved H2020 ERIGrid 2.0 project partners, especially Evangelos Kotsakis of the Joint Research Center (JRC) of the European Commission, for contributions to the specification and development of the test case profiles.

## Abbreviations

**CHIL**  Controller Hardware-in-the-Loop  
**FS**  Functional Scenario  
**HIL**  Hardware-in-the-Loop  
**HTD**  Holistic Test Description  
**ICT**  Information and Communication Technology  
**IED**  Intelligent Electronic Device  
**PHIL**  Power Hardware-in-the-Loop  
**RI**  Research Infrastructure  
**SIL**  Software-In-the-Loop  
**SuT**  System under Test  
**TC**  Test Case  
**TCP**  Test Case Profile  
**TRL**  Technology Readiness Level  
**UC**  Use Case